\documentclass[epjCONF,columns]{svjour} 
\usepackage{graphics,epsfig}
\usepackage[varg]{txfonts} 
\usepackage[latin1]{inputenc}

\usepackage{xspace}

\def\Journal#1#2#3#4{{#1} {\bf #2}, #3 (#4)}

\def\PLB{{Phys. Lett.}  B}
\def\PRL{Phys. Rev. Lett.}
\def\PRD{{Phys. Rev.} D}
\def\PR{Phys. Rep.}

\def\JHEP{J.\ High Energy Phys.}
\def\EPJC{Eur. Phys. J. C}

\def\met {\mbox{\ensuremath{\not\!\! E_T}}\xspace}

\newcommand{\GeV}{\ensuremath{\mathrm{Ge\kern-0.1em V}}\xspace}
\newcommand{\TeV}{\ensuremath{\mathrm{Te\kern-0.1em V}}\xspace}
\newcommand{\pt}{\ensuremath{p_{T}}\xspace}

\newcommand{\xo}{\mbox{\ensuremath{\tilde{\chi}^0_1}}}
\newcommand{\xpm}{\mbox{\ensuremath{\tilde{\chi}^{\pm}_1}}}

\renewenvironment{thebibliography}[1]
        {\begin{list}{\arabic{enumi}.}
        {\usecounter{enumi}\setlength{\parsep}{0pt}
         \setlength{\itemsep}{0pt} 
         \settowidth
        {\labelwidth}{#1.}\sloppy}}{\end{list}}

\session-title{HadronColliderPhysics Symposium 2011}
\begin{document}
%
%
\title{SUSY searches at the Tevatron}
\author{Michel Jaffr\'e\thanks{\email{jaffre@lal.in2p3.fr}}, on behalf of the CDF and D0 collaborations}
\institute{Laboratoire de l'Acc\'el\'erateur Lin\'eaire (Orsay, France)}
\abstract{
The Tevatron collider has provided the CDF and D0 collaborations with large datasets as input to a rich program of physics beyond the standard model.
The results presented here are from recent searches for SUSY particles using up to 6~fb$^{-1}$ of data. 
} 
\maketitle
\section{Introduction}
\label{intro}

Supersymmetry (SUSY)~\cite{susy} is one of the most favored theories beyond the standard model (SM).
Each SM particle is associated to a sparticle whose spin differs by one half unit.
This boson-fermion symmetry is obviously broken by some unknown mechanism. 
Even in the minimal supersymmetric extension of the SM (MSSM~\cite{mssm}) there are a large number of free parameters.
To reduce this number one can introduce new assumptions on the symmetry
breaking mechanism and build models based on minimal supergravity (as mSUGRA~\cite{msugra}) or
on a Gauge Mediated Symmetry Breaking scenario (GMSB~\cite{gmsb}), a top-down approach.
Another possibility is to make phenomelogical assumptions to reduce
the number of particles accessible to the experiment while keeping some of the
properties of the above models (bottom-up approach).

As the sparticles are heavy, to produce them one has to
make collisions at the highest center of mass energy.
The Tevatron was the best place for discovery until the start of LHC.
In the near term, Tevatron experiments and their large datasets remain competitive in areas like
production of third generation squarks and of non-coloured sparticles.  
 
I will report on recent results from the CDF and D0 collaborations,
assuming  R-parity~\cite{rparity} is conserved, i.e the sparticles are produced in pairs, and the lightest of
them (LSP) is stable, neutral, weakly interacting, and detected as missing transverse energy, \met.

\section{Scalar bottom and top quarks}
\label{sec:1}
In the MSSM, the mass splitting between the mass eigenstates of the two scalar partners
of a SM fermion depends on the mass of the fermion.
As such, the lightest scalar partners of the third generation may be light enough
to be produced copiously at the Tevatron.
In a data sample of 5.2~fb$^{-1}$, D0 has searched for a scalar bottom quark assuming
it decays exclusively into a bottom quark and the lightest neutralino (LSP), resulting in events with two $b$-jets and large \met.
This topology is identical to that for $p\bar{p}\to ZH \to \nu\bar{\nu}+b\bar{b}$ production, and the two 
analyses are based on the same trigger and event selection criteria~\cite{zhpaper}.
The SM background processes which contribute in this topology are the production of $W/Z$ bosons in association with $b$-jets and  top quark production.
No excess of events is observed above the expected SM processes which allows D0
to increase the excluded domain in the $(m_{\tilde{b}}, m_{\tilde{\chi}_1^0})$ mass plane
excluding a $247$~\GeV scalar bottom for a massless scalar neutralino and a $110$~\GeV neutralino for $160 < m_{\tilde{b}}< 200$~\GeV~\cite{d0-sbottom-2b+MET}.

\begin{figure}[htb]
  \begin{center}
   \resizebox{0.75\columnwidth}{!}{
    \includegraphics{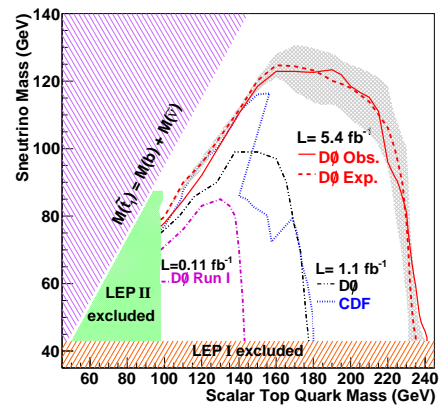}}
  \end{center} 
  \caption{
     D0 observed and expected 95\% C.L. exclusion contour in  the sneutrino and scalar top mass plot, and comparison with previous results.}
  \label{fig:d0-stop-limit}
\end{figure}

Scalar top quarks have been also searched for in various decay channels.
The most recent analysis is from D0 with a 5.4~fb$^{-1}$ data sample.
The scalar top quark is assumed to decay exclusively in the three body decay mode $\tilde{t} \to b l \tilde{\nu}$
with equal fraction to each lepton type, $l$\,; the scalar neutrino, $\tilde{\nu}$,  is either the LSP or decays invisibly to a $\nu$ and the LSP.
The event selection requires exactly one isolated electron and one isolated muon of opposite charge,
with transverse momenta, $\pt > 15$ and 10~\GeV respectively.
The SM backgrounds are from the Drell-Yan process ($\gamma/Z* \to \tau \tau$),
or from diboson and  top quark pair production. 
Several combinations of estimators are built to discriminate signal from the
different backgrounds depending on  the mass difference, $\Delta m = m_{\tilde{t}}-m_{\tilde{\nu}}$.
Scalar top masses below 210~\GeV are excluded for a scalar neutrino mass below 110~\GeV and $\Delta m >$~30~\GeV~\cite{d0-stop-2b+e+mu+MET}
(Fig.~\ref{fig:d0-stop-limit}).

\section{Diphotons and large \met}
\label{sec:2}
\begin{figure}[htb]
  \begin{center}
   \resizebox{0.75\columnwidth}{!}{
     \includegraphics{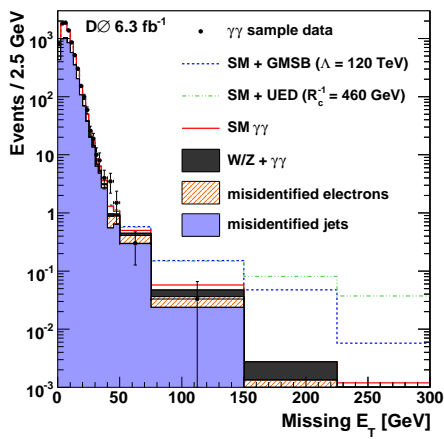}}
  \end{center} 
  \caption{D0 observed \met distribution in the diphoton event sample; expected distribution from SM backgrounds with and without contribution of GMSB events.}
  \label{fig:d0-gmsb-met}
\end{figure}

 Events with two high transverse momentum photons and large \met are relatively
rare in SM processes.
This topology is then very attractive for testing the GMSB model where the LSP
is the gravitino, a very light and weakly interacting particle.
In the case where the next-to-lightest SUSY particle (NLSP) is the neutralino, $\xo$, it will decay to a photon and a gravitino.
At the Tevatron, a large cross section is expected from chargino-neutralino pair
production which will cascade decay to two NLSP's and other leptons or jets, followed by the NLSP decays to a pair of photons and gravitinos.
D0 has analysed 6.3~fb$^{-1}$ of data~\cite{d0-GMSB-diphoton+MET} requiring events with two photon candidates with $\pt > 25$~\GeV. 
Events with real photons from SM processes and multijet events where jets are
faking one or more photons appear mostly at low \met, as shown in Fig.~\ref{fig:d0-gmsb-met}.
Signal is expected at large \met where no excess of events is observed.
D0 obtained a quantitative result when considering the Snow Mass Slope scenario, SPS8~\cite{sps8}; an effective SUSY breaking scale below 124~\TeV is excluded at 95\%C.L., as well as gaugino masses $m_{\tilde{\chi}_1^0}<175~\GeV$ and $m_{\tilde{\chi}_1^{\pm}} < 330~\GeV$.

\section{SUSY simplified models}
\label{sec:3}
Same charge dilepton events are a signature which has very low SM backgrounds,
essentially from $WZ$ and $ZZ$ diboson production. 
CDF has developped a general model independent strategy using simplified
 models as described in ref.~\cite{simplemod}.

\subsection{Same charge dileptons with jets}

\begin{figure}[t]
  \begin{center}
   \resizebox{0.75\columnwidth}{!}{
    \includegraphics{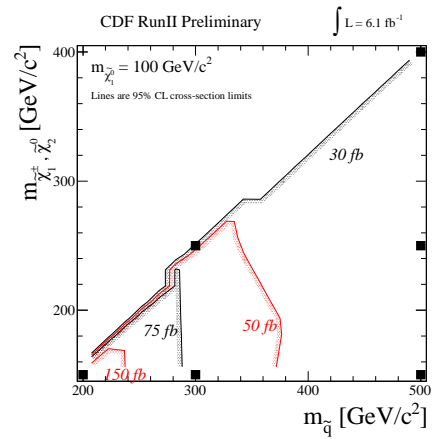}}
  \end{center} 
  \caption{CDF limits on squark pair production for a neutralino mass of 100~\GeV.}
  \label{fig:cdf-ssdileptons-susy}
\end{figure}

 The analysis requires two isolated leptons of the same\break charge with $\pt > 20$~\GeV and
at least  two jets with $\pt > 15$~\GeV.
The two leptons are either $ee$, $\mu\mu$ or $e\mu$.
This topology is  a SUSY signature; it results from  the decays of squark and gluino pairs.
The simplified model restricts  the decay chain 
to proceed only through $W$ or $Z$ bosons.
Sleptons are much heavier such that they have not influence on the event selection.
The model is further simplified by considering only the first generation of squarks 
and assuming that squarks decay equally to $\tilde{\chi}_1^{\pm}$ and $\tilde{\chi}_2^0$
which are degenerate in mass.
Finally the only parameters of the  model are the masses : $m_{\tilde{q}}$ ,$m_{\tilde{g}}$, 
$m_{\tilde{\chi}_1^{\pm}} = m_{\tilde{\chi}_2^0}$ and  $m_{\tilde{\chi}_1^0}$, the mass of the LSP.
The acceptance is largely affected by the mass difference between the ${\tilde{\chi}_1^{\pm}}$, ${\tilde{\chi}_2^0}$ and the LSP.
In a 6.1~fb$^{-1}$ data sample, CDF has not observed any significant deviations over the background expectations~\cite{cdf-ssdileptons+jets}.
Cross section limits on squark and gluino pair production are provided as a function of the sparticle masses.
In  Fig.~\ref{fig:cdf-ssdileptons-susy}, the limits are obtained for gluinos heavier  squarks and a 100~\GeV LSP.

\subsection{Same charge dileptons including a tau}
\begin{figure}[htb]
  \begin{center}
   \resizebox{0.75\columnwidth}{!}{
    \includegraphics{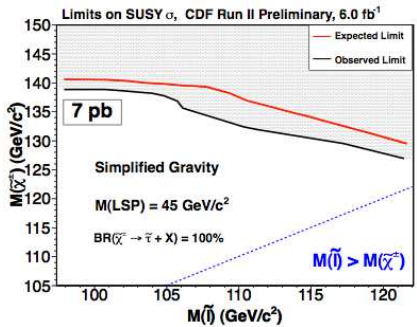}}
  \end{center} 
  \caption{CDF limits in a simplified gravity model where the LSP mass is 45~\GeV.}
  \label{fig:cdf-ssdileptons-simplifiedgravity-lsp45}
\end{figure}

In this analysis~\cite{cdf-sslepton+tauh}, CDF requires that one of the lepton is a $\tau$ decaying hadronically, the other one being either an $e$ or a $\mu$.
Two classes of  models are considered : a simplified gravity model reproducing some of the properties of mSUGRA and
a simplified  model motivated by gauge mediated SUSY breaking.
In both classes, \xpm and \xo are degenerate in mass and decay to sleptons.
The slepton decays to their SM partners and the LSP, either the $\tilde{\chi}_1^0$ or the gravitino.
In these decay chains, 2 cases were considered either all lepton flavors are equally probable or
modes involving $\tau$ are favored.
The multijet background from QCD jets faking a $\tau$ has been reduced by requiring a minimum value for $H_T$ defined as the
scalar sum of the $\tau$ \pt, lepton \pt and \met in the event.
Cross section limits are presented for each model in the ($m_{\tilde{\chi}_1^{\pm}}$,$m_{\tilde{l}}$) mass plane.
As an example, Fig.~\ref{fig:cdf-ssdileptons-simplifiedgravity-lsp45} shows the excluded domain
in the simplified gravity model for a 45~\GeV LSP and a $\xpm$ decaying exclusively to $\tilde{\tau} X$.

\section{Trileptons}
\label{sec:5}
\begin{figure}[htb]
  \begin{center}
   \resizebox{0.75\columnwidth}{!}{
    \includegraphics{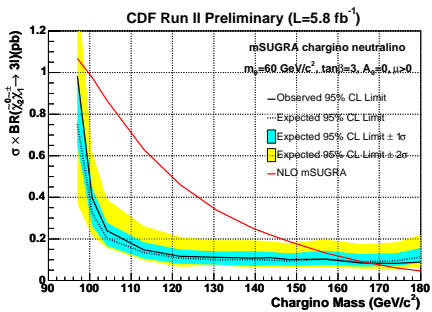}}
  \end{center} 
  \caption{CDF 95\% CL upper limit on the chargino-neutralino production cross
 section with subsequent trileptonic decay as a function of the chargino mass.}
  \label{fig:cdf-trilepton-limit}
\end{figure}

At the Tevatron, the trilepton final state is known as the 
``SUSY golden mode'' because of the small background level.
It is obtained by the production of a  $\tilde{\chi}_1^{\pm}$ and $\tilde{\chi}_2^0$ pair, which subsequently decay via $W$ and $Z$ bosons
(if sleptons are heavy) or via sleptons.
The first case suffers from the small leptonic branching ratios of the $W$/$Z$ decays.
In the second one, tau production may be enhanced if the lightest $\tilde{\tau}$ is much lighter than the other sleptons.
In a 5.8~fb$^{-1}$ data sample, CDF analysed events with two electrons or two muons requiring either 
the third lepton to be an identified lepton ( $e$, $\mu$ or $\tau$ decaying hadronically) or an isolated track~\cite{cdf-trileptons}.
In addition to a larger luminosity, this analysis benefits from  an extension of the lepton acceptance to the forward region,
and from a decrease of the minimum \pt of the non-leading leptons down to 5~\GeV.
A large effort was devoted to study the background yields in a lot of SM dominated control regions. 
Overall, no excess of events has been observed over the expectation leading to
a limit on the cross section times branching ratio into three leptons of 0.1~pb.
Interpreted within the mSUGRA model, this limit excludes a $\tilde{\chi}_1^{\pm}$
mass below 168~\GeV for a particular set of values for the other parameters ( Fig.~\ref{fig:cdf-trilepton-limit}).

\section{Leptonic jets}
\label{sec:6}
\begin{figure}[htb]
  \begin{center}
   \resizebox{0.75\columnwidth}{!}{
    \includegraphics{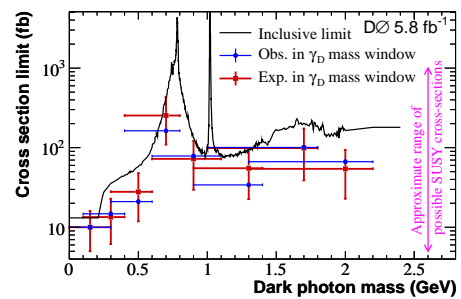}}
  \end{center} 
  \caption{Limit on the observed cross section for the three channels ($ee$, $\mu\mu$, $e\mu$) combined
    as a function of the dark photon mass.}
  \label{fig:d0-leptonic-limit}
\end{figure}

Hidden Valley (HV) scenarios~\cite{HV} introduce a hidden sector which is weakly coupled to SM particles.
They become popular as they provide a convincing explanation of observed astrophysical anomalies and discrepancies
in dark matter searches.
New low mass particles are introduced in the hidden sector, and the dark photon, which is the force carrier, would have a mass
 around 1~\GeV or less and would decay into a fermion or pion pair.
The case of decays to a lepton pair (electron or muon) is particularly attractive. 
SUSY is often included in HV models, and one could have a situation where the lightest neutralino will decay to a dark photon and $\tilde{X}$,
the lightest SUSY particle of the hidden sector, which will escape detection, leading to large \met.
As the dark photon is light, it will be highly boosted in the neutralino decay, and the two leptons will be close to each other.
Experimentally, one has to change the isolation criteria usually applied to identify leptons.
The presence of a track of opposite charge close to the lepton candidate will identify the so-called leptonic jet ($l$-jet).
Using 5.8~fb$^{-1}$ of data, D0~\cite{d0-leptonic-jets} has searched for pair production of $l$-jets in three configurations :
$ee$, $\mu\mu$ and $e\mu$.
No evidence of $l$-jets is observed in the distributions of the electron and muon pair masses.
Limits on the production cross section, around 100~fb for a 1~\GeV dark photon, are obtained ( Fig.~\ref{fig:d0-leptonic-limit}).
They are substantially weaker when the dark photon branching ratio to hadrons is larger, particularly near
the $\rho$ and $\phi$ resonances.

\section{Charged massive long lived particles (CMLLP)}
\label{sec:7}

\begin{figure}[htb]
  \begin{center}
   \resizebox{0.75\columnwidth}{!}{
    \includegraphics{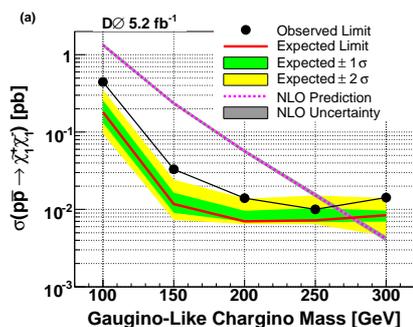}}
  \end{center} 
  \caption{D0 95\% C.L. cross-section limit as a function of the mass of a gaugino-like chargino.}
  \label{fig:d0-cmllp-g-chargino-limit}
\end{figure}

Some recent extensions of SM predict the existence of\break CMLLPs.
Indeed, they may help resolve difficulties of the model of the big bang 
nucleosynthesis at explaining the lithium abundance.
Within SUSY, charginos could be long lived if they are almost degenerate in mass with the LSP.
Experimentally, a long live time means that a CMLLP will decay outside the sensitive volume of the detector,
a large mass means a low speed and high ionisation loss.
CMLLPs would ressemble slow moving heavy muons.
In D0~\cite{d0-cmllp}, their low speed is measured by the time of flight recorded by the
muon scintillation counters; their energy loss compared to minimum ionising
muons is measured by the silicon detectors.
The analysis requires at least one well identified muon of high \pt.
Muons from meson decays are rejected by imposing isolation criteria in the tracking system and in the calorimeter.
Then, the remaining background originates from mismeasured real muons.
A background model has been built from data, essentially the production of $W$ bosons decaying
leptonically into a muon and a neutrino.
Requiring the $W$ transverse mass to be below 200~\GeV and a high speed for the muon  selects a signal free region.
The absence of any excess of events in 5.2~fb$^{-1}$ of data allows one to derive cross section limits of about 0.01~pb
for the production of a pair of CMLLPs with masses between 200 and 300~\GeV.
Comparing this limit to chargino pair production gives  mass limits on charginos
according to their nature,being  mostly a gaugino (Fig.~\ref{fig:d0-cmllp-g-chargino-limit}) or mostly a higssino.
The results of the analysis include also the limits for a light scalar top, after taking into account the complications
of hadronization and charge exchange of this strong interacting particle
between the production vertex and the muon system.

\section{Summary}
\label{sec:9}
Despite the start of LHC, SUSY searches at the Tevatron are still active.
With their large datasets, CDF and D0 have proved that there are domains
where they have competitive results.
Further details on physics results from the CDF and D0 collaborations can be obtained 
respectively from :
\begin{description}
\item http://www-cdf.fnal.gov/physics/physics.html
\item http://www-d0.fnal.gov/Run2Physics/WWW/results.htm
\end{description}

\section*{Acknowledgments}
The author would like to thank the CDF and D0 working groups for providing
the material for this talk.

\end{document}